\RequirePackage{fix-cm}
\documentclass[twocolumn]{svjour3}                     
\smartqed  
\usepackage{caption}
\usepackage{endnotes}
\usepackage{amsmath}
\usepackage{setspace} 
\usepackage[colorlinks=true,citecolor = blue,urlcolor=blue]{hyperref} \let\footnote=\endnote
\usepackage{bbm}
\usepackage{booktabs}
\def\ef#1{{\bf \color{gray}#1}}
\usepackage{enumitem}
\usepackage{amsfonts}
\usepackage{xcolor}
\definecolor{darkblue}{HTML}{0F07AD}
\usepackage{physics}
\usepackage{mathtools}
\usepackage{subfig} 
\usepackage[toc,page]{appendix}

\let\footnote=\endnote

\usepackage[numbers]{natbib}
 \def\bibfont{\small}%
 \def\bibsep{\smallskipamount}%
 \def\bibhang{24pt}%
\usepackage{graphicx}

\usepackage{url}

\usepackage{breakurl}
\journalname{Health Care Management Science}
\begin{document}
\sloppy

\title{Prediction of Personal Protective Equipment Use in Hospitals During COVID-19
}

\titlerunning{Prediction of PPE in Hospitals During COVID-19}        

\author{Eugene Furman \and Alex Cressman \and Saeha Shin \and Alexey Kuznetsov \and Fahad Razak \and Amol Verma \and Adam Diamant}

\institute{
        E. Furman \at
              Department of Operations Management, Rotman School of Management, Toronto, ON
          \and
          A. Cressman \at
              Division of General Internal Medicine, Temerty Faculty of Medicine, University of Toronto, Toronto, ON
            \and
            S. Shin \at
            Li Ka Shing Knowledge Institute, St. Michael’s Hospital, Toronto, ON
            \and
            A. Kuznetsov \at
            Department of Mathematics and Statistics, York University, Toronto, ON
            \and
            F. Razak, A. Verma, \at
            Division of General Internal Medicine and Li Ka Shing Knowledge Institute, St. Michael’s Hospital; 
            Department of Medicine and
            Institute of Health Policy, Management, and Evaluation, University of Toronto, Toronto, ON
          \and
            A. Diamant \at
              Department of Operations Management and Information Systems, Schulich School of Business, Toronto, ON
}

\date{Received: date / Accepted: date}

\maketitle

\begin{abstract}
Demand for Personal Protective Equipment (PPE) such as surgical masks, gloves, and gowns has increased significantly since the onset of the COVID-19 pandemic. In hospital settings, both medical staff and patients are required to wear PPE. As these facilities resume regular operations, staff will be required to wear PPE at all times while additional PPE will be mandated during medical procedures. This will put increased pressure on hospitals which have had problems predicting PPE usage and sourcing its supply. To meet this challenge, we propose an approach to predict demand for PPE. Specifically, we model the admission of patients to a medical department using multiple independent $M_t/G/\infty$ queues. Each queue represents a class of patients with similar treatment plans and hospital length-of-stay. By estimating the total workload of each class, we derive closed-form estimates for the expected amount of PPE required over a specified time horizon using current PPE guidelines. We apply our approach to a data set of 22,039 patients admitted to the general internal medicine department at St. Michael’s hospital in Toronto, Canada from April 2010 to November 2019. We find that gloves and surgical masks represent approximately 90\% of predicted PPE usage. We also find that while demand for gloves is driven entirely by patient-practitioner interactions, 86\% of the predicted demand for surgical masks can be attributed to the requirement that medical practitioners will need to wear them when not interacting with patients. 
\end{abstract}
\section*{List of Highlights}
\ef{\begin{enumerate}
    \item We propose an analytical demand prediction tool for PPE usage based on queueing theory that can be used in settings where PPE consumption is a function of a patient's length-of-stay in the hospital.
    \item The modeling approach is flexible; it can be deployed at multiple scales (departmental, hospital, regional), settings (outbreaks or regular operations), and can be used by administrative personnel for operational planning and inventory management.
    \item The general framework can accommodate the wide-variability in patient volumes between institutions, differences in the nature of typical patient-practitioner interactions at the ward-level, and distinct hospital policies governing default PPE usage in non-patient encounters (e.g., mandatory masking). Thus, it is expected to support planning activities for various health systems to ensure the equitable distribution of PPE during times of constrained supply.
    \item The analytical framework can be applied to non-medical settings, in particular, those where PPE consumption is a function of the length of a customers interaction with the organization.
\end{enumerate}}
\section{Introduction} \label{sec:introduction}
Personal protective equipment (PPE) includes items such as surgical masks, face shields,
gloves, eye protection, and gowns \citep{HControlsCA}. They are designed to
protect the wearer, and individuals they come in contact with, from potential
exposure to infectious diseases or other toxins \citep{PPEUS}. Although PPE is typically used in
clinical settings, it has become an essential commodity following the recent
outbreak of Coronavirus Disease (COVID-19). That is, to combat the spread of the
virus, many governments are mandating the use of PPE in public spaces such as retail stores, restaurants, community centers, and on public transit \citep[e.g.,][]{TransmissionOHealth}. Wearing a mask to conduct
activities outside the home is now recommended by the World Health Organization
\citep{who}, the Centers for Disease Control and Prevention \citep{newsroomcdc},
and the Government of Canada \citep{canada}. This non-pharmaceutical
intervention is designed to slow the spread of COVID-19, however, it has also
resulted in large surges in demand for PPE and, correspondingly, critical supply
shortages \citep{gondi2020personal}. This has had a detrimental effect on the
ability of hospitals to source PPE \citep{livingston2020sourcing} and outfit
their staff \citep{ranney2020critical}. In some cases, the inability to provide
adequate PPE to frontline health care workers has led to higher rates of infection and death amongst patients \citep{balmer2020italy}.

In hospitals, PPE has traditionally been used to protect healthcare workers when
performing various types of medical procedures \citep{akduman1999use,
benson2013proper}. During the pandemic, however, PPE has become a requirement for virtually all patient-practitioner interactions; any time a health worker enters a patient's room or physically interacts with a patient, they may be required to wear PPE. 
As a result, although patient volumes initially decreased with the onset of the pandemic as many non-emergent procedures were postponed, there has been a large increase in the use of PPE to manage urgent and non-elective patient care \Citep{hfma}. For instance, in
response to the COVID-19 pandemic, the Canadian government has ordered approximately 395 and 154 million surgical and N95 masks, respectively, to distribute directly
to hospitals \citep{GCAN}. As acute care facilities resume normal operations
(e.g., diagnostic testing, elective surgery, ambulatory care), all staff,
employees, and visitors will likely be required to wear PPE at all times
\Citep{uhn2020} while additional PPE requirements will be mandated during
medical procedures \Citep{OHealth}. 
This will put even more pressure on PPE
supply chains which, in some health care systems, face estimated delays of up to
6 months and have had major distributors unable to fill orders
\Citep{mehrotra2020personal}. Since one of the biggest obstacles to restarting
normal hospital operations is the consistent and timely supply of PPE, these
statistics are particularly troubling \Citep{hfma}.

Given the importance of PPE in acute care centers, proactive PPE management has become an essential component in hospital operations \citep{CBC}. Successful
administration of PPE inventory is directly linked to accurately predicting the demand for medical services, and in particular, the number and nature of all patient-practitioner interactions \citep[see][for instance]{ppeCovid20}. Doing
so is challenging due to the large number of diagnoses, clinical procedures, and surgical interventions as well as the time-dependent nature of patient arrivals
\citep[e.g.,][]{yom2014erlang}. While various simulation studies have been used to estimate hospital workload during the pandemic
\citep{calafiore2020modified,toda2020susceptible,wangping2020extended}, they are hard to replicate, time-consuming to build, difficult to use effectively, and are not conducive to performing a comparative analysis that is required for prescriptive managerial decision-making.

In this work, we develop a time-varying queueing model to predict the amount of PPE required in a clinical inpatient setting over a specified time horizon. As has been well-established in the literature \citep[e.g.,][]{whitt2017data, yom2014erlang}, we assume that the process governing when patients arrive to the hospital is time-dependent. We then cluster patients with similar hospital experiences (e.g., investigations, interventions) into classes and estimate their length-of-stay (LoS) in the hospital as well as the PPE requirements for each interaction with a practitioner. 
We show that these dynamics can be modeled using
multiple independent $M_t/G/\infty$ queues \citep[see][for instance]{massey1993networks}, one for each patient class, and as a consequence, derive closed-form estimates for the expected amount of PPE required during the
time horizon.


Using a large data set of clinical, demographic, and operational attributes from 22,039 patients admitted to the general internal medicine (GIM) service at St. Michael's Hospital (a primary care facility in Toronto, Canada) from April 2010 to November 2019, we demonstrate the practicality of our approach. We first validate the assumption that time-varying demand is an appropriate modelling choice. We then describe how to group patients into classes depending on the nature of their medical interactions as well as their LoS values. Note that this is an important step to ensure that patients in the same class have similar hospital experiences. Next, we use our model to predict the yearly PPE requirements of the GIM service at St. Michael's Hospital when it returns to normal operations excluding those patients who are diagnosed with COVID-19. 
Using the current regulations governing PPE use at the hospital and leveraging pre-pandemic patient volumes, we show that the GIM service will need approximately 225,000 gloves, 11,500 gowns, 181,500 surgical masks, 7500 N95 masks, and 4000 face shields. Thus, gloves and surgical masks represent approximately 90\% of the predicted PPE usage. We also find that while demand for gloves is driven entirely by patient-practitioner interactions, 86\% of the predicted demand for surgical masks can be attributed to the requirement that medical practitioners will need to wear masks when not interacting with patients.
In addition, we show that our approach provides upper and lower bounds for the amount of PPE that is predicted to be used. We also perform an analysis to determine the sensitivity of the predictions to the number of patient classes chosen by the modeller. 


We contribute to the operations research and medical literature by applying a queueing theoretical framework to a high-impact medical problem. To the best of our knowledge, our work is the first to obtain closed-form expressions for PPE usage in a hospital setting. Our method is analytical, computationally
efficient, and does not require that a hospital develop an extensive simulation study. By deriving closed-form
expressions, the sensitivity of the predictions to changes
in the model's parameters can be evaluated. This helps hospital administrators gain practical insight into the dynamics of PPE usage which is especially valuable for the effective management of a scarce resource in a rapidly changing environment. Finally, we note that our approach is easily scalable; it can be used to make predictions for a single department, an entire hospital, or be deployed at the regional or provincial level.

\section{Literature Review and Contribution} \label{sec:literatureReview}
To predict PPE consumption, we introduce a stochastic queueing framework with
multiple independent $M_t/G/\infty$ queues to model the dynamics of distinct
patient classes that are admitted to the hospital, receive clinical care, and
interact with practitioners. Pioneering theoretical work in the study of
$M_t/G/\infty$ systems date back to \citet{palm1943variation} and
\citet{khintchine1955mathematical} who show that the number of jobs in the system
at any time instant
follows a Poisson process
with a time-varying rate. Since then, the extant literature has shown that
departures from such queues also follows a non-homogeneous Poisson process
\citep[see][]{brown1969invariance,foley1982non,foley1986stationary}. More recent work derives the expected number of jobs remaining in the system
after each departure, i.e., the number of busy servers, for specific service
distributions \cite{eick1993mt,eick1993physics}. Further,  several studies derive the steady-state distribution and fluid limit of systems with a periodic arrival rate
\cite{dong2015stochastic,dong2015using,whitt2015many}. For a review of queueing systems with non-stationary demand, see
the survey papers by \citet{defraeye2016staffing} and \citet{whitt2017time}.

From a practical perspective, the number of applications that use
$M_t/G/\infty$ queues to model service systems is vast: they have been
employed, for instance, to evaluate the adequacy of storage systems
\citep{crawford1977wrsk}, determine the readiness of military equipment
\citep{hillestad1980models, crawford1981palm}, estimate the occurrence of bugs
in software testing \citep{yang1998infinite}, and model the arrival of customers
to in-bound call centers \citep{khudyakov2010designing,vizarreta2018assessing}.
Specific to healthcare, several studies have used the model to analyse
practitioner staffing and capacity management problems
\citep{yom2014erlang,pender2016risk, furman2019customer,razak2020modelling}.
Due to the assumption of infinite capacity,
$M_t/G/\infty$ queues are particularly useful in situations where service delay
is near zero \cite{green1998note}. The principle of zero waiting time is common in the estimation of
total workload for staffing analyses and is also known as the offered load
approximation \citep{feldman2008,janssen2011, liu2018staffing, furman2020cloud}.
For instance, \citet{vericourt2011nurse} model a medical unit as a closed queueing network and determine optimal nurse-to-patient ratios. There are also several papers that analyze the supply of hospital beds and derive expressions to promote better management strategies in settings with
time-varying demand \cite{green2001strategies,green2007,zeltyn2011simulation,zychlinski2018time}. We contribute to this literature by using multiple
$M_t/G/\infty$ queues to derive closed-form expressions to predict PPE consumption from an offered load estimate of hospital workload.

Our work is related to the literature that develops best-practices for
supply chain disruptions. \citet{tang2006robust,stecke2009sources} and
\citet{carbonara2018real} provide insight into how a supply chain can respond to
natural disasters, terrorist attacks, and other unforeseeable emergencies.
Logistics networks can be built with redundant transportation routes
\citep{dash2013emerging}, suppliers are encouraged to invest in more robust
infrastructure \citep{dolinskaya2018humanitarian}, and inventory postponement
can be employed to better understand the changing demand-supply relationship
\citep{chiou2002adoption,yeung2007postponement,choi2012postponement}.
Nevertheless, especially in demand-driven supply chains, these approaches are
not always useful in situations with extreme demand volatility unrelated to
infrastructure damage or logistical disturbances 
\citep{chan2004multi,chen2009demand, verdouw2011framework}.
Instead, effective inventory management and accurate demand predictions are
crucial \citep{chen2001coordination, milner2002flexible,qi2004supply,xu2003demand}. We add to this literature by proposing an analytical demand
prediction tool for PPE usage that can be employed in settings with supply chain disruptions where consumption is a function of the length of a customers interaction with an organization.

Specific to research on COVID-19, our analysis is related to studies that
predict future demand for medical services; see the surveys by
\citet{sahin20202019,workman2020endonasal} and \citet{harapan2020coronavirus}. Since the
onset of the pandemic, this literature has grown substantially. Some studies
employ deterministic compartmental modifications of
Susceptible-Infected-Recovered (SIR) models which are parameterized by empirical
studies \citep{tuite2020mathematical,calafiore2020modified,biswas2020covid}.
Such methods result in systems of differential equations that must be solved
numerically to obtain predictions or insight related to possible public health initiatives \citep{liu2020reproductive}. Other studies 
combine dynamic SIR models with Bayesian inference techniques \citep[see][for example]{chen2020scenario} or propose stochastic
Markov models to predict the spread of the disease \citep[see][for example]{zhang2020prediction}; solutions are obtained by
performing a simulation analysis. Stochastic implementations of
SIR models are also common in the literature \citep{bardina2020stochastic,
karako2020analysis,simha2020simple}. 
We provide an approach that can be
used alongside these models. In particular, given a PPE policy and a (potentially) time-varying demand curve for hospital services using one of the above methods, our model derives a closed-form expression for PPE usage and can be employed during a COVID-19 outbreak or after regular operations have resumed.


Finally, our work contributes to the literature on critical shortages of PPE
during the COVID-19 pandemic. While many studies leverage COVID-19 transmission
models to evaluate the effectiveness of non-pharmaceutical containment
strategies \citep[e.g.,][]{evgeniou2020epidemic,flaxman2020estimating,zhang2020optimal},
the literature predicting demand for PPE is scarce. Some authors propose
qualitative techniques to manage PPE in a medical setting
\citep{rowan2020challenges,ranney2020critical}. These strategies are consistent
with practices that are used when there are demand and/or supply disruptions in
the pharmaceutical industry \citep[see][for example]{fox2009ashp}. Other papers
use simulation-based frameworks to derive PPE usage \citep{ppeCovid20}. These
approaches are difficult to reproduce, and thus, their estimation error is hard
to quantify. Our work is the first to propose an analytical predictive model of PPE demand in a clinical setting that can be deployed at multiple scales (departmental, hospital, regional), settings (outbreaks or regular
operations), and can also be independently used by administrative personnel for operational planning and supply management.

\section{Model Formulation and Workload Estimation}\label{sec:model}
In this section, we introduce a general stochastic queueing model and describe its suitability in estimating the amount of PPE required for a hospital department. Let $\mathcal{I}$ be a set of patient classes defined using managerially-relevant features, for instance, demographic characteristics, patients with varying acuity levels, clinical diagnoses, and length-of-stay. Classes should be chosen such that all patients in class-$i\in\mathcal{I}$ have similar care paths, i.e., a sequence of medical investigations and interventions, and LoS values.
Class $i$ patients are assumed to arrive to the hospital and be admitted
according to a non-homogeneous Poisson process $\Lambda_i(t)$ with time-varying
intensity \ef{$\lambda_i(t)\in\mathbb{C}^1$}.
Further, each class-$i$ patient stays at the hospital for a random time $S_i$ which represents their length-of-stay (LoS); we define the corresponding stochastic vector $\boldsymbol{S}\coloneqq (S_1,S_2,...,S_I)$. The LoS for each patient within each class is independent and identically distributed where class-$i$ patients have cumulative distribution function $G_i$. 
Finally, we assume that $S_i$ is independent 
from $\Lambda_i(t)$ for any time $t\in \mathbb{R}$.

Our goal is to estimate the total clinical workload of a hospital department,
which in turn, will allow us to predict the PPE required. 
\ef{In practice, a decision to admit a patient to an inpatient service is only possible if there is available capacity. In general, hospital capacity is continuously monitored and dynamically adjusted to meet regional demand; only in the most extreme circumstances will patients be turned away  \citep{OHACapPlan, OHA}.}
Consequently, we do not restrict hospital capacity and instead, assume that practitioners can provide medical care to any admitted patient as soon as they arrive. That is, we estimate the total workload of a hospital department by aggregating the workload from $I=|\mathcal{I}|$ independent $M_t/G/\infty$ queues leveraging the
merging/splitting property of a Poisson process. \ef{In this regard,
our approach provides a theoretical upper bound on PPE usage. However, this modelling assumption has been used in many service settings including those that model demand for hospital resources  \citep{palomo2020flattening}. Furthermore, it represents a good approximation of reality since our study of PPE consumption begins after a patient is physically admitted to the hospital and usage is based on historical data which implies that all arriving patients were served.}

Inferring the workload from such systems is a standard modelling technique in the operations literature 
\citep{eick1993physics,massey1993networks,feldman2008}. In addition,
patients transferred from one clinical service to another
are considered discharged by the former and newly admitted by the latter. 
Such events are rare and thus, we can consider these individuals as new arrivals for estimation purposes. Note that the intensive care unit (ICU) constitutes an exception to this rule: between 5 to 10 percent of GIM patients are transferred to the ICU at least once over the duration of their treatment. In this study, we consider the ICU as an external service, and, thus, subtract the times patients spend there from their total length-of-stay.


Let $\{\Delta_i(t)|t \in \mathbb{R}\}$
be a headcount stochastic process corresponding to the number of class $i$ patients being
discharged over the interval $[0,t]$.
Applying Theorem~1 in \citet{eick1993physics}, we
obtain the steady-state probability distribution of $\Delta_i(t)$. Because the GIM service has been continuously operating for a long time, the steady-state assumption is appropriate in our setting.  
Specifically, the number of class $i$ patients discharged over the interval $[0,t]$ is given by $\{\Delta_i(t)|t \in \mathbb{R}\}$ which is a non-homogeneous Poisson process with mean
\begin{align}\label{deltatExp}
\mathbb{E}[\Delta_i(t)] &\coloneqq \int_0^t\int_{0}^{\infty} \lambda_i(u-s)dG_i(s)du &\forall i\in\mathcal{I}.
\end{align}
Notice that, following the framework of \citet{eick1993physics}, we assume $t\in\mathbb{R}$ but only consider the dynamics of the system at times $t\ge 0$. 

Unfortunately, for most LoS distributions, \eqref{deltatExp} must be computed numerically as closed-form expressions do not exist unless, for example, $G_i$ is exponentially distributed. In addition, the departure process $\Delta_i(T)$ is dependent on the LoS of class-$i$ patients. As a result, we condition on the individual quantiles of the LoS distribution for each class $i \in \mathcal{I}$. More specifically, let $\sigma_i$ be the desired quantile value for class-$i$ patients where we define $\pmb{\sigma}\coloneqq(\sigma_1,\ldots,\sigma_I)$ and let $\Delta_i(t;\sigma_i)$ denote the departure process of class-$i$ patients conditioned on $S_i = \sigma_i$ for each $i \in \mathcal{I}$. Thus, $\{\Delta_i(t;\sigma_i) |t \in \mathbb{R}\}$ is a headcount stochastic process that represents the number of class $i$ patients discharged over the interval $[0,t]$ with LoS value equal to $\sigma_i$. This corresponds to a non-homogeneous Poisson process with mean
\begin{align}\label{deltatExpConditional}
\mathbb{E}[\Delta_i(t,\sigma_i)] &\coloneqq \int_{0}^{t} \lambda_i(u-\sigma_i)du &\forall i\in\mathcal{I}.
\end{align}





\subsection{Prediction of Demand for PPE}

Multiple types of PPE are used in clinical settings, such as surgical masks, N95 respirators, gloves, face shields, etc. Further, demand for different kinds of
PPE varies depending on the nature of the interaction between patients and practitioners as well as current public health regulations and institutional guidelines \citep[see][for instance]{OHealth,ppeCovid20}. Thus, we assume that a hospital uses $N$ different types of PPE in its daily operations.

Total demand of PPE comprises all protective equipment used by employees, i.e.,
medical staff, and patients.
Although, in this study, we assume that patients admitted to the hospital occupy separate rooms and do not need to wear PPE while on their own, our model can be naturally extended to account for patients with shared accommodations. Further, hospital policy dictates that clinicians wear a surgical mask and a face shield for all interactions with hospitalized patients. Additional precautions may be used by hospital staff and clinicians when performing particular procedures and/or assessments. There may also be separate regulations for patients who are placed in a higher level of isolation, such as those diagnosed with COVID-19.
As a result, we define $Q^m_n$ to be the total quantity of type $n\in\{1,2,...,N\}$ PPE used by employees when no interaction with patients takes place and $Q^u_{i,n}$ to be the amount of type $n$ PPE used by medical staff during interactions with class $i$ patients.
Thus, the total demand for type $n$ PPE is given by
\begin{align}\label{ttlDemand}
Q_n & :=  Q^m_n + \sum_{i=1}^I Q^u_{i,n} &\forall n\in\{1,2,\ldots,N\}.
\end{align}
We assume, without loss of generality, that PPE is not reused but discuss this extension in Section~\ref{ppeEstimationResults}.

Define $\boldsymbol{m} \coloneqq (m_1,m_2,\ldots,m_N)^{\prime}$ to be a vector such that element $m_n$ represents the average number of type $n$ PPE items used daily by
an employee when not interacting with patients. Then,
\begin{align}
Q^m_n &=  m_n  W(T) &\forall n\in\{1,2,\ldots,N\},
\end{align} where $W(T)$ is the number of estimated work days of all medical employees over the planning horizon. In particular, note that we assume $Q^m_n$ increases linearly in the workload $W(T)$. Discussions with medical practitioners indicate
that this is the most appropriate model.

Suppose there are $J$ different categories of clinical interactions such as nursing (e.g., vital signs measurement, medication administration), physician visits, medical testing, and surgical procedures. 
Define an $I\times J$ matrix $\boldsymbol{C}$ where element $c_{i,j}$ is the average daily number of clinical interactions from category $j$ that are required by a class $i$ patient (note: median values can also be used to reduce the effect of outliers although we did not observe any appreciable difference in our results). We also define an $I\times J$ matrix $\boldsymbol{U}_n$ such that element $u_{i,j}^n$ represents the average number of type $n$ PPE items used during each category $j$ interaction with a patient of class $i$ (see Table~\ref{modelPara} in the Appendix). \ef{Then, conditioning on the LoS value $\boldsymbol{S}=\boldsymbol{\sigma}$ as well as the average daily number of clinical interactions and the corresponding PPE usage, we estimate $Q^u_{i,n}$ to be}
\begin{align}
\hat{Q}^u_{i,n} &= \sigma_i\Delta_i(T;\sigma_i)\sum_{j=1}^Jc_{i,j}u_{i,j}^n \\
& \qquad \qquad \forall i\in\{1,2,\ldots,I\},\forall n\in\{1,2,\ldots,N\}, \nonumber
\end{align}

Notice that $c_{i,j}u_{i,j}^n$ represents the average daily number of type $n$ PPE used by class $i$ patients during all medical interactions belonging to category $j$. Aggregating over each $j$ and multiplying by the stochastic quantity $S_i$ gives the average number of type $n$ PPE used by a class $i$ patient during their length-of-stay in the hospital. Finally, multiplying these terms by the integral of the headcount stochastic process gives the average amount of type $n$ PPE used by all class $i$ patients discharged over the specified time horizon $T$.


As noted above, $\Delta_i(T)$ and $S_i$ are dependent, i.e., the number of discharged patients at time $t$ is a function of the LoS of class-$i$ patients. This makes deriving the marginal expectation of $Q_n$ cumbersome to obtain. Instead, in the following lemma, we leverage \eqref{deltatExpConditional} and derive the conditional expectation of $Q_n$ given that the LoS of class-$i$ patients is fixed to a given quantile.
\begin{lemma}[Conditional Expectation]\label{condQn}
For every $i$, suppose $\sigma_i>0$ and $T>\sigma_i$. Then,
\begin{align}\label{condEst}
\mathbb{E}[\hat{Q}_n]&= \sum_{i=1}^I  \sigma_i\sum_{j=1}^Jc_{i,j}u_{i,j}^n\int_{0}^{T} \lambda_i(u-\sigma_i) du\\ &+m_nW(T),\textit{ }
\forall n\in\{1,2,\ldots,N\}.\notag
\end{align}
\end{lemma}
Equation \eqref{condEst} is derived by conditioning on a particular quantile of the LoS distribution. For example, if $\sigma_i = \mathbb{E}[S_i]$ for all $i \in \mathcal{I}$, then for a class-$i$ patient, \eqref{condEst} considers the dynamics of the average stochastic path of the departure process $\Delta_i(t;\sigma_i)$ as the total number of paths grows to infinity. Further, as the variances of the hospital LoS and the average daily counts of medical interactions decrease, the gap between the conditional \ef{$(\hat{Q}_n)$} and unconditional expectation of the demand for type $n$ PPE ($Q_n$) also decreases. Thus, \eqref{condEst} provides a better approximation to the demand for PPE if the classes of patients are selected such that their LoS and treatment requirements are relatively similar; this motivates why patients should first be clustered into $I$ classes.

\begin{table*}[h]
\centering
\caption{Summary of the notation.}
\label{table:nota}
\scalebox{0.9}[0.90]{
\begin{tabular}{ll}
  \toprule
$\lambda_i(t)$ & class $i$ patient's rate of admission to the hospital\\
$S_i$ & random variable corresponding to the hospital length-of-stay of a class $i$ patient\\
$\boldsymbol{S}$ & $I\times 1$ stochastic vector of length-of-stay random variables\\
$\sigma_i$ & desired quantile value chosen for the length-of-stay distribution of a class $i$ patient\\
$\pmb{\sigma}$ & $I\times 1$ vector of  length-of-stay quantile values\\
$G_i(t)$ & cumulative distribution function for the length of stay of class $i$ patient\\
$\Delta_i(t)$ & stochastic process counting the number of class $i$ patients discharged over $[0,t]$\\
$\Delta_i(t;\sigma_i)$ & stochastic process counting the number of class $i$ patients discharged over $[0,t]$ conditional on $S_i=\sigma_i$\\
$Q_n$ & total stochastic demand for type $n$ PPE\\
$Q^m_n$ & total stochastic demand for type $n$ PPE by all hospital employees while not interacting with patients\\
$Q^u_{i,n}$ & total stochastic demand for type $n$ PPE by all hospital employees during their interactions with class $i$ patients\\
$\boldsymbol{C}$ & $I\times J$ matrix of average daily counts of medical interactions $j$ required by class $i$ patient\\
$\boldsymbol{m}$ & $N\times 1$ vector of average daily counts of type $n$ PPE used by hospital employees while not interacting with patients\\
$\boldsymbol{U}_n$ & $I\times J$ matrix of average number of type $n$ PPE used during medical interaction $j$ with a class $i$ patient\\
   \bottomrule
\end{tabular}
}
\end{table*}
\section{Data Description and Results}\label{sec:num}
We apply our approach to estimate the PPE needs for the GIM service at St. Michael's hospital. The GIM accounts for approximately 40\% of all emergency department admissions to the hospital \cite{verma2017patient} and cares for patients with a broad range of diseases \cite{verma2018prevalence} while focusing on cases with complex medical needs.  
Because the operations at St. Michael's Hospital is directly affected by the COVID-19 pandemic, effective prediction of PPE usage is critical to their inventory planning and their ability to deliver adequate medical care. 

To parameterize our predictive model, we used 9 years of data from April 2010 to November 2019 collected from St. Michael's Hospital by the General Medicine Inpatient Initiative (GEMINI) \cite{verma2017patient}. The data set includes both administrative and clinical records of discharged patients. GEMINI data sets have been rigorously validated and are demonstrated to be highly reliable \cite{pasricha2020assessing}.
Our data set comprises of 37,492 hospital admissions for 22,039 unique patients whose median age is 66 years old (52, 79), where values in brackets correspond to the first and third quartile, respectively.
Approximately 43\% of hospital admissions to the GIM are by female patients and the five most common clinical diagnoses are chronic obstructive pulmonary disease and bronchiectasis (6\%), pneumonia (5\%), acute cerebrovascular disease (5\%), urinary tract infections (5\%), and gastrointestinal hemorrhages (4\%).

The median value for LoS is 4.83 days (2.58, 9.54)
which suggests an asymmetrical probability distribution. We determine the average daily counts of medical interactions per patient (\ef{see  Table~\ref{CMatPara} in the Appendix for an example of the average counts of each interaction per patient type}) as well as the corresponding type of interaction and PPE usage from the data set and by interviewing medical experts in the partner hospital (see the Appendix for details on the semi-structured interview protocol). Notice that Table~\ref{modelPara}, provided in the Appendix, displays the average amount of PPE used during all medical interactions in addition to the items already worn by clinical staff when not interacting with patients. Thus, in cases where no additional PPE is required, the value in the table is equal to zero. Alternatively, some medical interactions are conducted by multiple practitioners which means that a larger amount of PPE is required. For example, surgical procedures typically require two porters, a surgeon and one or two trainees, an anesthesiologist, and two nurses. 

When not interacting with patients, medical staff require two surgical masks per shift (which is approximately 12 hours in length) and one face shield per week. The GIM service at St. Michael's Hospital requires 50 nurses, 4 phlebotomists, 10 porters, 20 doctors, 3 physiotherapists, 3 occupational therapists, 2 dietitians, 2 language pathologists, and 3 discharge planners each day. For simplicity, we assume that shifts of all medical staff are of the same length. Notice that this assumption is easy to relax. Finally, we consider seven types of PPE ($N=7$): gloves, gowns, surgical masks, N95 masks, face shields, bouffants, and boot covers.

We use equation \eqref{condEst} to derive an annual estimate of PPE usage by clustering all patients into  classes based on the nature of their medical interactions as well as their length-of-stay within the hospital. 
To account for the aforementioned asymmetry in the LoS distribution, and since \eqref{condEst} computes a conditional expectation, we evaluate PPE usage assuming that LoS remains at one of its quantile values for each class.
We fix our planning horizon ($T$) to one year (365 days)
and estimate the value of $\int_{0}^{T-\sigma_i} \lambda_i(u)du$ by calculating the number of class-$i$ discharges that occur during a typical year prior to the pandemic. In the remainder of this section, we confirm that a non-homogeneous Poisson
distribution best describes the arrival process.
We then discuss how we cluster patients into classes and present estimates of the projected annual PPE usage.

\subsection{Testing the Non-homogeneous Poisson Assumption}
Because our data set contains the arrival times and discharge times of each patient, the number of discharges from the GIM over a planning horizon can be computed without evaluating the integral in \eqref{condEst}. However, there are many cases where such fine-grained data is not available. In such settings, only arrival times and/or LoS values may be accessible. In other cases, the prediction interval set by the modeller may be sufficiently short (e.g., daily or weekly) which necessitates the evaluation of a functional form of departing patients at time $t$. 
In these scenarios, computing the integral is essential. Therefore, both for completeness and to ensure that the analytical representation of the demand for PPE in \eqref{condEst} is valid, we test the assumption that the arrival process follows a non-homogeneous Poisson distribution.

We closely follow the procedure described in \citet{brown2005statistical},
i.e., we test the null hypothesis ($H_0$) that admissions to the GIM follow a Poisson distribution with a piecewise constant rate. To do this, we break up the planning horizon into progressively smaller non-overlapping time intervals. Note that, for this analysis, we consider admissions to the GIM from the two most recent years in order to account for possible changes in the demand for GIM services. We then continue to decrease the length of these intervals until the arrival rate remains stationary over at least 90\% of the constructed intervals. We test the hypothesis of stationarity by applying the
Kolmogorov-Smirnov (KS) test and confirming that, for each time interval, the logarithmically transformed arrival times can be modeled by independent standard exponential
random variables.

\begin{table}[h]
\centering
\caption{Testing the non-homogeneous Poisson assumption for different time intervals.}\label{nhpTest}
\scalebox{0.8}{
\begin{tabular}{c|c|c}
\toprule
  Number of Intervals & Length (days) & \%  Not Rejected By KS Test \\ 
  \midrule
10                  & 90.8                         & 0.00                        \\ 
20                  & 43.0                         & 35.00                       \\ 
30                  & 28.2                          & 63.30                     \\ 
40                  & 21.0                         & 80.00                       \\ 
80                  & 10.3                         & 88.75                    \\ 
800                 & 1.00                             & 90.38                    \\ 
\bottomrule
\end{tabular}
}
\end{table}
According to Table~\ref{nhpTest}, as the length of each interval reduces to one day, the arrival rates over 90\% of the intervals follow a Poisson distribution with a stationary rate according to the KS test (0.05 significance level). This implies that a 
non-homogeneous Poisson distribution best describes the arrival rate and that a $M_t/G/\infty$ modelling framework is appropriate for this application.
 
 \subsection{Clustering Results}
To ensure patient classes have similar care paths and LoS values, we cluster patients into groups based on the nature of their medical interactions (15 types) and LoS (see Table~\ref{modelPara} in the appendix). In our data set, each medical intervention is captured by a set of timestamps. \ef{To avoid counting the same patient-practitioner interaction multiple times, as advised by our medical co-authors, we assume that all timestamps within a one-hour interval are related to a single interaction.} This assumption is critical as some interactions between patients and practitioners result in multiple timestamps that are minutes apart (e.g., vital signs, the administration of drugs, and laboratory test collections) and, thus, reflect a single episode of PPE use. \ef{Robustness checks confirm that while varying this time interval only slightly affects the estimates, reducing the interval to zero is susceptible to a significant amount of double-counting.}

\begin{figure}[h]
         \caption{Clustering results.}
              \subfloat[Cluster Visualization]{%
        \includegraphics[height=7.0cm,width=0.45\textwidth]{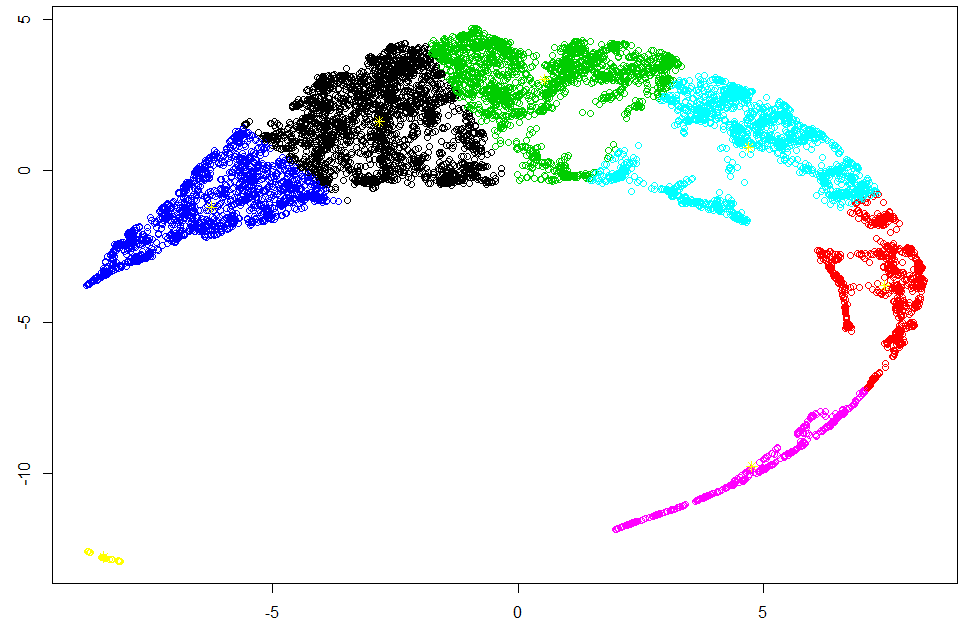}%
        \label{fiveClust}%
         }
         \hfill
     \subfloat[Elbow Plot]{%
        \includegraphics[height=8.5cm,width=0.45\textwidth]{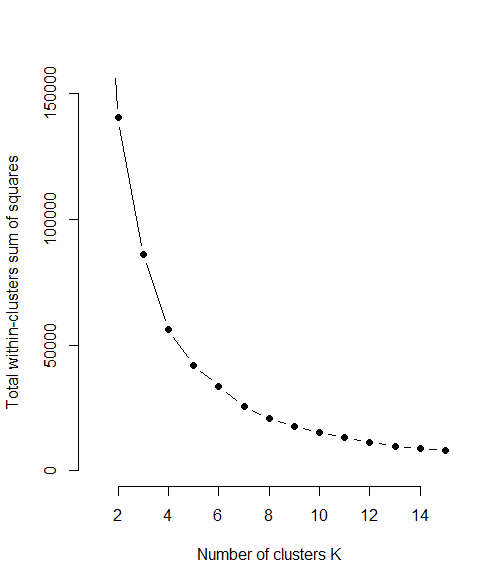}%
        \label{elbowPlot}
         }
\end{figure}

We use the Uniform Manifold Approximation and Projection (UMAP) algorithm paired with the k-means clustering algorithm to group patients into classes. UMAP is a dimensionality reduction technique based on Riemannian geometry and algebraic topology that projects high-dimensional data (15 types of medical interactions and LoS) onto a two-dimensional space; see Figure~\ref{fiveClust} for a visual representation. The smaller total squared error within a cluster implies that there is a high similarity of patients assigned to that class. It also improves the quality of our conditional estimate of demand for PPE. Thus, we determine the optimal number of clusters by applying the k-means clustering algorithm which minimizes the total squared error within each cluster. We then use the elbow method to determine the best value of $k$ \cite{joshi2013modified}; \ef{this balances the tradeoff between minimizing the within-cluster variance while ensuring that the number of patients in each cluster is large enough such that the corresponding LoS estimate is meaningful.}

As demonstrated in Figure~\ref{elbowPlot}, the within cluster error decreases slowly as the number of clusters exceeds 7; adding more clusters does not model the data significantly better. 
For more information on the clustering approach, please see the Appendix and Table~\ref{errorWithin}.

To illustrate the effect of our clustering procedure, we present a quantile summary of the LoS (days) distribution by comparing non-clustered patients to the clustered results. \ef{Having relatively similar LoS values in each cluster is important as we would like its within-cluster variation to be small so that the gap between our conditional estimates and their corresponding marginal quantities are negligible.}
\begin{table}[h]
\caption{The effect of clustering on the different quantiles for the LoS distribution (days).}\label{effectLoS}
\centering
\scalebox{0.9}{
\begin{tabular}{lccccc}
\toprule
                      & 0\%  & 25\% & 50\% & 75\%  & 100\% \\ \midrule
Cluster 1 of 1 ($100\%$)       & 0.0    & 1.9  & 3.9  & 7.9   & 354.2 \\ \midrule
Cluster 1 of 7 ($18\%$) & 0.0  & 0.5  & 0.8  & 1.4   & 4.8 \\
Cluster 2 of 7 ($27\%$) & 0.1  & 1.7  & 2.3  & 2.9   & 6.4 \\
Cluster 3 of 7 ($22\%$) & 0.4  & 3.7  & 4.5  & 5.2   & 7.1 \\
Cluster 4 of 7 ($17\%$) & 5.3  & 6.9  & 7.9  & 9.3   & 11.7 \\
Cluster 5 of 7 ($10\%$)  & 10.8 & 12.7 & 14.3 & 16.6  & 20.9 \\
Cluster 6 of 7 ($6\%$) & 20.4 & 24.2 & 29.2 & 35.8  & 57.0 \\
Cluster 7 of 7 ($1$\%) & 59.0 & 65.6 & 82.0 & 128.2 & 354.2 \\ \bottomrule
\end{tabular}
}
\end{table}
If all patients are assigned to a single class, one quarter stay in the GIM between 0 and 1.9 days (first quartile); similarly, 25\% of patients remain in the GIM more than 7.9 but less than 354.2 days (fourth quartile). The clustered patients, however, have more similar LoS ranges.
In particular, cluster one contains patients who remain in the GIM for a very short period of time, cluster two and three are assigned patients who stay in the hospital less than one week, cluster four includes patients who stay in care less than 11 days, cluster five includes patients with LoS shorter than 20 days, and cluster six includes patients who stay in the facility significantly longer. Cluster seven, which contains approximately 1\% of patients, represent the departments heaviest users. We note that some clusters have overlapping LoS ranges because other factors describing their care path differ. 

\subsection{PPE Estimation Results} \label{ppeEstimationResults}
We apply equation \eqref{condEst} to compute the total demand for type $n$ PPE using 5, 6, 7, and 8 cluster partitions. To describe its distribution, we condition our estimates 
on the quartiles of the LoS and present the results in Table~\ref{results},
where the first and third rows per each cluster quantity correspond to the lower and
upper bounds of PPE usage.
\begin{table*}[h]
\centering
\caption{Prediction of PPE usage as a function of the number of clusters.}\label{results}
\scalebox{1.0}{
\begin{tabular}{@{}cccccccc@{}}
\toprule
LoS Quartile & Gloves  & Gowns  & Surgical Masks & N95 Masks & Face Shields & Bouffants & Boot Covers \\ \midrule
\multicolumn{8}{c}{Five Clusters  ($I=5$)}                                                                    \\ \midrule
Q1           & 122,771 & 6,422  & 169,193       & 4,094     & 3,906        & 6,422     & 6,422       \\
Median       & 206,459 & 10,748 & 180,093       & 6,891     & 3,906        & 10,748    & 10,748      \\
Q3           & 264,107 & 13,785 & 187,208       & 8,787     & 3,906        & 13,785    & 13,785      \\ \midrule
\multicolumn{8}{c}{Six Clusters  ($I=6$)}                                                                     \\ \midrule
Q1           & 134,232 & 6,935  & 169,917       & 4,385     & 3906       & 6,935     & 6,935       \\
Median       & 219,111 & 11,348 & 180,954       & 7,239     & 3906       & 11,348    & 11,348      \\
Q3           & 279,440 & 14,517 & 188,221       & 9,203     & 3906       & 14,517    & 14,517      \\ \midrule
\multicolumn{8}{c}{Seven Clusters  ($I=7$)}                                                                   \\ \midrule
Q1           & 129,216 & 6,779  & 169,233       & 4,229     & 3,906        & 6,779     & 6,779       \\
Median       & 226,007 & 11,721 & 181,774       & 7,476     & 3,906        & 11,721    & 11,721      \\
Q3           & 277,995 & 14,433 & 187,989       & 9,161     & 3,906        & 14,433    & 14,433      \\ \midrule
\multicolumn{8}{c}{Eight Clusters ($I=8$)}                                                                   \\ \midrule
Q1           & 151,878 & 7,839  & 171,964       & 4,980     & 3,906         & 7,839     & 7,839       \\
Median       & 229,751 & 11,850 & 182,296       & 7,610     & 3,906         & 11,850    & 11,850      \\
Q3           & 274,123 & 14,163 & 187,491       & 9,051     & 3,906         & 14,163    & 14,163      \\ \bottomrule
\end{tabular}
}
\end{table*}
According to the seven-cluster estimates in Table~\ref{results}, 
gloves and surgical masks are the most prevalent items as they constitute 
90\% of the total PPE predicted.
Further, the annual usage of gowns represents only 3\% (similarly to bouffants and boot covers) of the total (454,324) amount of PPE used, while N95 masks constitute only 2\%. As a reminder, due to the nature of our data, these estimates account for non-COVID-19 patients only, i.e., those patients who are not under investigation for the Coronavirus. However, our model is flexible enough and can accommodate these patients as separate classes if the data becomes available.

Table~\ref{results} 
also helps to understand the sensitivity of our results to the number of patient classes specified by the modeller. 
In general, we observe higher predictions in the amount of PPE as the number of clusters increases. This is because the average and median values of features included in the clustering procedure are more heavily influenced by larger-valued observations. However, the increase in predicted PPE usage with the number of clusters is sample specific; data sets with fewer outliers may have a decreasing pattern. 
Although using more clusters decreases
the total squared error, fewer data points contribute to the length-of-stay estimate. This may lead to an inaccurate prediction for the LoS distribution even though patients may have similar care plans. Furthermore, the estimates may overfit to the data in the sample.
Thus, we advise that a modeller does not increase the number of clusters too far beyond the point that is recommended by the elbow method. 

We find that some types of PPE, such as surgical masks and face shields, show little variation in forecasted demand. That is, their quartile estimates are similar regardless of the number of patient classes chosen. This is because the majority of the annual need for these types of PPE occur when practitioners are not interacting with patients; the estimate is 156,220 and 3,906 for surgical masks and face shields, respectively. Thus, while demand for gloves is solely driven by the number of medical interactions, 86\% of surgical mask use is driven by the requirement that medical employees must wear a mask whilst in the hospital.

Finally, we note that the above approach can be adapted to address situations where PPE can be reused. In particular, let $\gamma_n$ be the proportion of type $n$ PPE which can be reused over $r_n$ interactions. Then, \ef{$(1-\gamma_n)\mathbb{E}[\hat{Q}_n]+\frac{\gamma_n}{r_n}\mathbb{E}[\hat{Q}_n]$} represents the total predicted demand of type $n$ PPE.

 \section{Conclusions}\label{sec:discConc}
In this paper, we use theory from time-varying queueing models to present a prediction framework that can be used to forecast the amount of PPE required over a specified time horizon. To this end, we first cluster patients with similar hospital experiences into classes and estimate their LoS in the hospital as well as the PPE requirements of each patient-practitioner interaction. By demonstrating that the dynamics of each patient class can be modelled using an $M_t/G/\infty$ queue, we present closed-form estimates for the expected amount of PPE required for each patient class and aggregate the results together to generate a prediction of PPE usage.

We contribute to the pandemic and supply chain disruption literature
by helping practitioners mitigate unexpected changes in demand when disruptions do not affect the operation of a service, but instead, prompt new mandatory regulations that affect the equipment used in its performance.
Moreover, our analysis provides bounded estimates that anticipate the time-variability in the system. In particular, using current PPE-usage guidelines under COVID-19, we find that 
the general internal medicine department at our partner hospital must anticipate much higher demand for gloves and surgical masks than gowns. The former comprises 90\% of the total 454,324 items predicted while the latter accounts for only 3\% of the annual PPE usage. In addition, our analysis suggests that only 14\% of demand for surgical masks in a hospital setting is caused by interactions with patients. Thus, an annual estimate of usage for this type of PPE is expected to be less volatile than the anticipated demand for gloves.
 
As suggested in Section~\ref{sec:model}, our approach is versatile and
computationally efficient. A simple application of Lemma~\ref{condQn} admits a back-of-the-envelope calculation.
In this case, the aggregate number of departures from the system per patient type as well as a quantile estimate for the LoS are sufficient to derive bounded conditional estimates of PPE usage. Contrary to \citet{ppeCovid20}, for instance, our predictions do not require
that an extensive simulation study be constructed; the technique we develop is not restricted to estimates of PPE during a quarantine and can be applied to other settings such as normal hospital operations. In addition, our approach may be used for a comparative analysis. For example, if patient classes are pre-specified by a medical practitioner, the demand for PPE can be estimated and compared for multiple choices of arrival functions and LoS distributions over a planning horizon of arbitrary length. Our time-varying queueing framework naturally accommodates this exploratory approach by providing an analytical way of estimating the total number of departures conditioned on a carefully selected LoS value.     

Although our PPE prediction tool can be applied to a wide variety of clinical settings, our study includes a number of data-specific limitations. \ef{First, our modeling approach assumes that the total hospital capacity is always sufficient to satisfy the demand. Thus, our analytical framework provides a theoretical upper bound on PPE usage. In practice, however, hospital records only contain information on patients who have been admitted. Thus, data-driven PPE usage predictions all implicitly make this assumption.}
Second, the guidelines governing the use of PPE for each type of medical interaction, as summarized in Table~\ref{modelPara} in the Appendix, is distinct to St. Michael's Hospital. These estimates may vary depending on the location and clinical focus of the medical institution under consideration. Third, we estimate the clinical workload generated by typical medical interactions based on the data collected prior to the COVID-19 pandemic, i.e., we exclude both confirmed COVID-19 patients and patients who are under investigation for the virus. As this data becomes available, the PPE needs for these patient categories can be estimated and added to the prediction model. Fourth, while 15 important types of clinical interactions are captured in the data set, some are represented more crudely than others. For example, a nurse who assists a patient with toileting or bathing is not captured. As a result, our approach may underestimate the hospital's overall PPE needs. However, these limitations may be addressed by collecting additional data. To this end, future research should seek to validate the predicted estimates of PPE usage against real-world demand.

Despite these limitations, our methodology complements ongoing efforts that help to manage supply chains during the COVID-19 pandemic.
For instance, using an arrival function estimated by SIR models,
we can derive the corresponding PPE requirements
over a planning horizon of arbitrary length. Our study also shows good synergy with emerging platforms that connect PPE suppliers to consumers \citep{PPEConnect, fsConnect} as consumers can more accurately predict their PPE usage and liaise with suppliers that have the requisite capacity.

\begin{acknowledgements}
The authors would like to acknowledge the important contributions of our key stakeholders including Shirley Bell (Nurse Manager for General Internal Medicine), David Nelson (Medical Imaging Manager), Janice Barnett (Cardiac Sonography and Team Leader), Dr. Samir Grover (Gastroenterologist), Dr. Kieran McIntyre (Respirologist) and Dr. Karim Ladha (Anaesthesiologist). 
\end{acknowledgements}

%
%

\bibliographystyle{spbasic}      
\bibliography{references}  

\section*{Appendices and Proofs of Statements}

\subsection*{Proof of Lemma~\ref{condQn}}
For $\sigma_i >0$ and $T>\sigma_i$ for all $i$, we apply the conditional expectation operator to \eqref{ttlDemand}. Then, given \eqref{deltatExpConditional}, and by the linearity of expectation, \eqref{condEst} holds.\hfill \qed

\section*{Clustering Procedure}\label{clusterAppend}
We cluster patients based on 16 variables which include a patient's length-of-stay  in the hospital (days) and their average daily count of 15 medical interactions as per Column 1 in Table~\ref{modelPara}. For example, a patient with a length-of-stay equal to 3.5 days and 12 vital sign measurements will be assigned an average daily count of $12/3.5=3.4$ measurements of vital signs.

To improve the quality of our clustering procedure, we employ Uniform Manifold Approximation and Projection technique (UMAP) \citep{mcinnes2018umap} as a 
pre-processing step. Contrary to other 
non-linear projection methods (t-SNE or Isomap, for instance), it does not favor the preservation of local distances over global distance. That is, using UMAP as a pre-processing step for clustering preserves both the local (dissimilarities within clusters) and global (dissimilarities between clusters) structure of the data set. Further, the algorithm is less computationally intensive than t-SNE, for instance, and in contrast to linear projection techniques such as Principal Component Analysis (PCA), does not attempt to construct multidimensional vectors to recreate the location of each data point. Thus, it is not vulnerable to 20\% - 30\% loss in representative accuracy.

Because we do not aim to predict cluster membership for future patients and would like to cluster all existing patients without exceptions, we choose a standard k-means approach. To ensure the stability of the clusters, we initialize the procedure with 25 random starting partitions (see nstart option for the kmeans function in the R documentation). Contrary to our needs, density based techniques, (HDBSCAN, for instance) may consider some of the data points as noise. \ef{Note that other techniques can be considered as long as they result in clusters with adequate similarity (i.e., the variance in the LoS values within a cluster are sufficiently small).} 

\begin{table}[h]
\centering
\caption{Within cluster variation as a function of the number of clusters used.}\label{errorWithin}
\scalebox{0.85}{
\begin{tabular}{cc}
\toprule
Number of Clusters, k & Total Squared Error Within Clusters \\ \midrule
1                     & 290,250                             \\
2                     & 140,491.9                           \\
3                     & 86,060.95                           \\
4                     & 56,175.68                           \\
5                     & 42,133.06                           \\ 
6                     & 33,761.11                           \\ 
7                     & 25535.31                           \\
8                     & 20880.25                           \\
9                     & 17563.34                           \\
10                    & 15222.13                            \\\bottomrule
\end{tabular}
}
\end{table}
\begin{table*}[ht]
\caption{Average PPE usage per patient-practitioner interaction.}\label{modelPara}
\centering
\scalebox{0.65}{
\begin{tabular}{|c|c|c|c|c|c|c|c|}
\hline
Interaction Types, $j$    & Gowns, $u_{1,j}$ & Gloves, $u_{2,j}$ & Surgical Masks, $u_{3,j}$ & N95 Masks, $u_{4,j}$ & Shields, $u_{5,j}$ & Bouffants, $u_{6,j}$ & Boot Covers, $u_{7,j}$ \\ \hline
Vital signs measurement   & 0                & 1                 & 0                        & 0                    & 0                  & 0                    & 0                      \\
Medication administration & 0                & 1                 & 0                        & 0                    & 0                  & 0                    & 0                      \\
Lab Test Collection                 & 0                & 1                 & 0                        & 0                    & 0                  & 0                    & 0                      \\
X-ray                     & 0                & 2                 & 0                        & 0                    & 0                  & 0                    & 0                      \\
CT                        & 0                & 2                 & 0                        & 0                    & 0                  & 0                    & 0                      \\
MRI                       & 0                & 2                 & 0                        & 0                    & 0                  & 0                    & 0                      \\
Ultrasound                & 0                & 2                 & 0                        & 0                    & 0                  & 0                    & 0                      \\
Nuclear Medicine          & 0                & 1                 & 0                        & 0                    & 0                  & 0                    & 0                      \\
Interventional Radiology  & 3.5              & 3.5               & 0                        & 3.5                  & 0                  & 3.5                  & 3.5                    \\
Transthoracic Echocardiography (TTE)                       & 0                & 1                 & 0                        & 0                    & 0                  & 0                    & 0                      \\
Transesophageal Echocardiography (TEE)                      & 3                & 3                 & 3                        & 3                    & 0                  & 3                    & 3                      \\
Bronchoscopy              & 4                & 4                 & 4                        & 4                    & 0                  & 4                    & 4                      \\
Dialysis                  & 0                & 1                 & 0                        & 0                    & 0                  & 0                    & 0                      \\
Surgical Procedure       & 5.5              & 5.5               & 4                        & 2                    & 0                  & 5.5                  & 5.5                    \\
Room Transfer             & 0                & 1.5               & 0                        & 0                    & 0                  & 0                    & 0                      \\ \hline
\end{tabular} 
}
\end{table*}

\begin{table*}[ht]
\caption{Average counts of interactions per patient type for the eight-cluster analysis.}\label{CMatPara}
\centering
\begin{tabular}{|c|c|c|c|c|c|c|c|c|}
\hline
Interaction Types, $j$                 & $c_{1,j}$ & \multicolumn{1}{l|}{$c_{2,j}$} & \multicolumn{1}{l|}{$c_{3,j}$} & \multicolumn{1}{l|}{$c_{4,j}$} & \multicolumn{1}{l|}{$c_{5,j}$} & \multicolumn{1}{l|}{$c_{6,j}$} & \multicolumn{1}{l|}{$c_{7,j}$} & \multicolumn{1}{l|}{$c_{8, j}$} \\ \hline
Vital signs measurement                & 3.51      & 4.70                           & 3.47                           & 3.47                           & 3.63                           & 3.73                           & 3.12                           & 3.53                            \\
Medication administration              & 1.95      & 15.83                          & 1.22                           & 2.77                           & 2.24                           & 4.64                           & 0.66                           & 1.51                            \\
Lab Test Collection                    & 2.33      & 12.79                          & 1.99                           & 2.59                           & 2.67                           & 4.74                           & 1.22                           & 2.06                            \\
X-ray                                  & 0.19      & 2.27                           & 0.14                           & 0.31                           & 0.23                           & 0.47                           & 0.07                           & 0.16                            \\
CT                                     & 0.10      & 1.05                           & 0.05                           & 0.21                           & 0.12                           & 0.27                           & 0.03                           & 0.07                            \\
MRI                                    & 0.03      & 0.04                           & 0.02                           & 0.07                           & 0.03                           & 0.05                           & 0.01                           & 0.02                            \\
Ultrasound                             & 0.05      & 0.27                           & 0.02                           & 0.06                           & 0.05                           & 0.09                           & 0.01                           & 0.03                            \\
Nuclear Medicine                       & 0.00      & 0.00                           & 0.00                           & 0.01                           & 0.01                           & 0.01                           & 0.00                           & 0.00                            \\
Interventional Radiology               & 0.02      & 0.01                           & 0.02                           & 0.01                           & 0.01                           & 0.01                           & 0.01                           & 0.02                            \\
Transthoracic Echocardiography (TTE)   & 0.03      & 0.06                           & 0.01                           & 0.05                           & 0.03                           & 0.05                           & 0.01                           & 0.02                            \\
Transesophageal Echocardiography (TEE) & 0.00      & 0.00                           & 0.00                           & 0.00                           & 0.00                           & 0.00                           & 0.00                           & 0.00                            \\
Bronchoscopy                           & 0.04      & 0.10                           & 0.02                           & 0.08                           & 0.05                           & 0.09                           & 0.01                           & 0.03                            \\
Dialysis                               & 0.01      & 0.04                           & 0.00                           & 0.01                           & 0.01                           & 0.01                           & 0.00                           & 0.00                            \\
Surgical Procedure                     & 0.04      & 0.45                           & 0.05                           & 0.05                           & 0.04                           & 0.05                           & 0.03                           & 0.05                            \\
Room Transfer                          & 0.18      & 1.06                           & 0.10                           & 0.32                           & 0.22                           & 0.43                           & 0.05                           & 0.14                            \\ \hline
\end{tabular}
\end{table*}
We use the total squared error within clusters as a single aggregate measure of similarity amongst patients. This is because a sample variance estimate is dependent on the size of the cluster. Table~\ref{errorWithin} presents the within cluster variation as a function of the number of clusters that are used. Due to the multi-dimensional nature of the data, the clustering technique aims to reduce the variation amongst all variables at the same time rather than focusing on one of them specifically. 
\section*{Estimation of the Number of Medical Interactions Per Day as well as PPE Usage}\label{PPEpara}

The first column in Table~\ref{modelPara} represents the most common types of medical interactions between patients and practitioners for individuals admitted to the GIM service at St. Michael's hospital. 
In columns 2-8, we display the count of type $n\in\{1,2,\ldots,7\}$ PPE used
during each type of interaction. While the values in column 1 are obtained by analyzing the types of interactions in the data set, the values in columns 2-8
were obtained by conducting semi-structured interviews with various medical practitioners in each sub-speciality of our partner hospital and summarizing their responses. 

More specifically, we engaged key stakeholders from clinical departments throughout St. Michael's Hospital. They included a nurse manager for the general internal medicine (GIM) department, a medical imaging manager, an echocardiography team leader and a cardiac sonographer, a dialysis charge nurse, a gastroenterologist, a respirologist, and an anesthesiologist. The semi-structured interviews were conducted with each individual and process-mapping techniques were used to understand the workflow, number of patient interactions, personnel needs, and the PPE usage per episode of patient care. For elective and non-elective surgeries, we used common surgical procedures conducted on GIM patients, including laparoscopic intra-abdominal surgeries or vascular procedures such as amputation, to inform the model. These interviews provided pragmatic estimates of PPE usage and helped to estimate the number of patient interactions on a daily basis.

\ef{For all patients that belongs to a cluster, we compute the average number of medical interactions as the total number of interactions divided by their corresponding LoS. Then, elements of $I\times J$ matrix $\boldsymbol{C}$ are arithmetic averages of these estimates over all patients of the same type. In columns 2-9 of Table~\ref{CMatPara}, we present these averages for each of the eight clusters (i.e., $c_{1,j}, \ldots, c_{8,j}$) and for every interaction type $j$ rounded to two decimal digits. Notice that there are some interactions which are rarely repeated - instances of bronchoscopy or surgical procedures have small values - while more frequent interactions (vital signs measurement, medicine administration, and lab test collection) admit larger values.}

\end{document}